\documentclass[aps,prl,nofootinbib,twocolumn,showpacs,superscriptaddress,letterpaper,amsmath,amssymb]{revtex4}
\usepackage{graphicx}  
\usepackage{dcolumn}   
\usepackage{bm}        
\usepackage{amssymb}   
\usepackage{feynmf}
\usepackage{slashed}
\usepackage{multirow}
\usepackage{url}
\usepackage{hyperref}
\usepackage{color}
\usepackage{array}

\newcolumntype{L}[1]{>{\raggedright\let\newline\\\arraybackslash\hspace{0pt}}m{#1}}
\newcolumntype{C}[1]{>{\centering\let\newline\\\arraybackslash\hspace{0pt}}m{#1}}
\newcolumntype{R}[1]{>{\raggedleft\let\newline\\\arraybackslash\hspace{0pt}}m{#1}}

\def\gsim{\lower0.5ex\hbox{$\:\buildrel >\over\sim\:$}}
\def\lsim{\lower0.5ex\hbox{$\:\buildrel <\over\sim\:$}}

\def \n{\noindent}

\let\n=\nu
\let\t=\tau

\newcommand{\be}{\begin{equation}}
\newcommand{\ee}{\end{equation}}
\newcommand{\bea}{\begin{eqnarray}}
\newcommand{\eea}{\end{eqnarray}}

\newcommand{\nbox}{{\,\lower0.9pt\vbox{\hrule \hbox{\vrule height 0.2 cm
\hskip 0.2 cm \vrule height 0.2 cm}\hrule}\,}}

\def\sub#1{_{\lower.25ex\hbox{$\scriptstyle#1$}}}

\def\slash{\not\!}

\newskip\zatskip \zatskip=0pt plus0pt minus0pt
\def\matth{\mathsurround=0pt}
\def\lsim{\mathrel{\mathpalette\atversim<}}
\def\gsim{\mathrel{\mathpalette\atversim>}}
\def\sigv{\ifmmode \langle\sigma v\rangle\else $\langle\sigma v\rangle$\fi}
\newskip\zatskip \zatskip=0pt plus0pt minus0pt
\def\matth{\mathsurround=0pt}
\def\lsim{\mathrel{\mathpalette\atversim<}}
\def\gsim{\mathrel{\mathpalette\atversim>}}
\def\atversim#1#2{\lower0.7ex\vbox{\baselineskip\zatskip\lineskip\zatskip
  \lineskiplimit
  0pt\ialign{$\matth#1\hfil##\hfil$\crcr#2\crcr\sim\crcr}}}

\begin{document}

\thispagestyle{empty}
\vspace*{-3.5cm}

\vspace{0.5in}

\title{Parameterized Machine Learning for High-Energy Physics}

\begin{center}
\begin{abstract}
We investigate a new structure for machine learning classifiers applied to problems in high-energy physics by expanding the inputs to include not only measured features but also physics parameters.  The physics parameters represent a smoothly varying learning task, and the resulting parameterized classifier can smoothly interpolate between them and replace sets of classifiers trained at individual values. This simplifies the training process and gives improved performance at intermediate values, even for complex problems requiring deep learning. Applications include tools parameterized in terms of theoretical model parameters, such as the mass of a  particle, which allow for a single network to provide improved discrimination across a range of masses. This concept is simple to implement and allows for optimized interpolatable results.
\end{abstract}
\end{center}

\author{Pierre Baldi}
\affiliation{Department of Computer Science, University of
  California, Irvine, CA 92697}
\author{Kyle Cranmer}
\affiliation{Department of Physics, NYU, New York, NY }
\author{Taylor Faucett}
\affiliation{Department of Physics and Astronomy, University of
  California, Irvine, CA 92697}
\author{Peter Sadowski}
\affiliation{Department of Computer Science, University of
  California, Irvine, CA 92697}
\author{Daniel Whiteson}
\affiliation{Department of Physics and Astronomy, University of
  California, Irvine, CA 92697}

\date{\today}

\pacs{}
\maketitle


\section{Introduction}

Neural networks have been applied to a wide variety of problems in high-energy physics~\cite{Denby:1987rk,Peterson:1993nk}, from event classification~\cite{Abreu:1992jp,Kolanoski:1995zn} to object reconstruction~\cite{Peterson:1988gs,Aad:2014yva} and triggering~\cite{Lonnblad:1990bi,Denby:1990wb}. Typically, however, these networks are applied to solve a specific isolated problem, even when this problem is part of a set of closely related problems. An illustrative example is the signal-background classification problem for a particle with a range of possible masses. The classification tasks at different masses are related, but distinct. Current approaches require the training of a set of isolated networks~\cite{Aaltonen:2012qt, Chatrchyan:2012tx}, each of which are ignorant of the larger context and lack the ability to smoothly interpolate, or the use of a single signal sample in training~\cite{Aad:2014xea,Chatrchyan:2012yca}, sacrificing performance at other values.

In this paper, we describe the application of the ideas in Ref.~\cite{cranmer2015} to a new neural network strategy, a {\it parameterized neural network}  in which a single network tackles the full set of related tasks. This is done  by simply extending the list of input features to include not only the traditional set of event-level features but also one or more {\it parameters} that describe the larger scope of the problem such as a new particle's mass.  The approach can be applied to any classification algorithm; however, neural networks provide a smooth interpolation, while tree-based methods may not.

A single parameterized network can replace a set of individual networks trained for specific cases, as well as smoothly interpolate to cases where it has not been trained. In the case of a search for a hypothetical new particle, this greatly simplifies the task -- by requiring only one network -- as well as making the results more powerful -- by allowing them to be interpolated between specific values. In addition, they may outperform isolated networks by generalizing from the full parameter-dependent dataset.  

In the following, we describe the network structure needed to apply a single parameterized network to a set of smoothly related problems and  demonstrate the application for theoretical model parameters (such as new particle masses) in a set of examples of increasing complexity.

\section{Network Structure \& Training}

\begin{figure}
\includegraphics[width=0.4\textwidth]{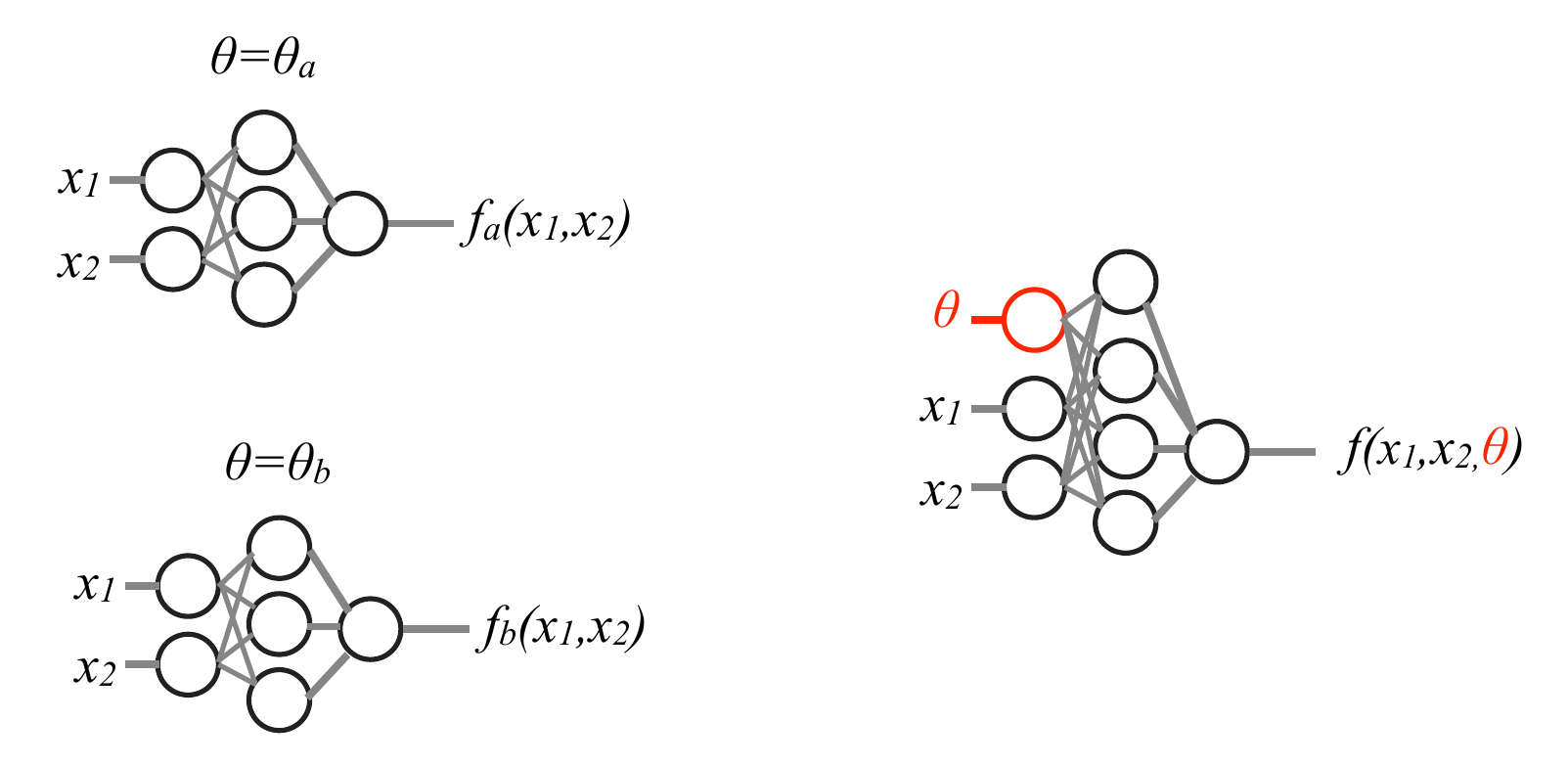}
\caption{ Left, individual networks  with input features $(x_1,x_2)$, each trained with examples with a single value of some parameter $\theta=\theta_a,\theta_b$. The individual networks are purely functions of the input features. Performance for intermediate values of $\theta$ is not optimal nor does it necessarily vary smoothly between the networks.  Right, a single network trained with input features  $(x_1,x_2)$ as well as an input parameter $\theta$; such a network is trained with examples at several values of the parameter $\theta$. }
\label{fig:networks}
\end{figure}

A typical network takes as input a vector of features, $\bar{x}$, where the features are based on event-level quantities. After training, the resulting network is then a function of these features, $f(\bar{x})$.  In the case that the task at hand is part of a larger context, described by one or more parameters, $\bar{\theta}$. It is straightforward to construct a network that uses both sets of inputs, $\bar{x}$ and $\bar{\theta}$, and operates as a function of both: $f(\bar{x},\bar{\theta})$.   For a given set of inputs $\bar{x}_0$, a traditional network evaluates to a real number $f(\bar{x}_0)$. A parameterized network, however, provides a result that is parameterized in terms of $\bar{\theta}$: $f(\bar{x}_0,\bar{\theta})$, yielding different output values for different choices of the parameters $\bar{\theta}$; see Fig.~\ref{fig:networks}. 

Training data for the parameterized network has the form $(\bar{x}, \bar{\theta}, y)_i$, where $y$ is a label for the target class. 
The addition of $\bar\theta$ introduces additional considerations in the training procedure. While traditionally the training only requires the conditional distribution of $\bar{x}$ given  $\bar{\theta}$ (which is predicted by the theory and detector simulation), now the training data has some implicit prior distribution over $\bar\theta$ as well (which is arbitrary).  When the network is used in practice it will be to predict $y$ conditional on both $\bar{x}$ and $\bar{\theta}$, so the  distribution of $\bar\theta$ used for training is only relevant in how it affects the quality of the resulting parameterized network  -- it does not imply that the resulting inference is Bayesian.  In the studies presented below, we simply use equal sized samples for a few discrete values of $\bar{\theta}$.   Another issue is that some or all of the components of $\bar{\theta}$ may not be meaningful for a particular target class. For instance, the mass of a new particle is not meaningful for the background training examples.  In what follows, we randomly assign values to those components of $\bar\theta$ according to the same distribution used for the signal class. In the examples studied below the networks have enough generalization capacity and the training sets are large enough that the resulting parameterized classifier performs well without any tuning of the training procedure. However, the robustness of the resulting parameterized classifier to the implicit distribution of $\bar\theta$ in the training sample will in general depend on the generalization capacity of the classifier, the number of training examples, the physics encoded in the distributions $p(\bar{x} | \bar{\theta}, y)$, and how much those distributions change with $\bar\theta$.

\section{Toy Example}

\begin{figure}
\includegraphics[width=0.35\textwidth]{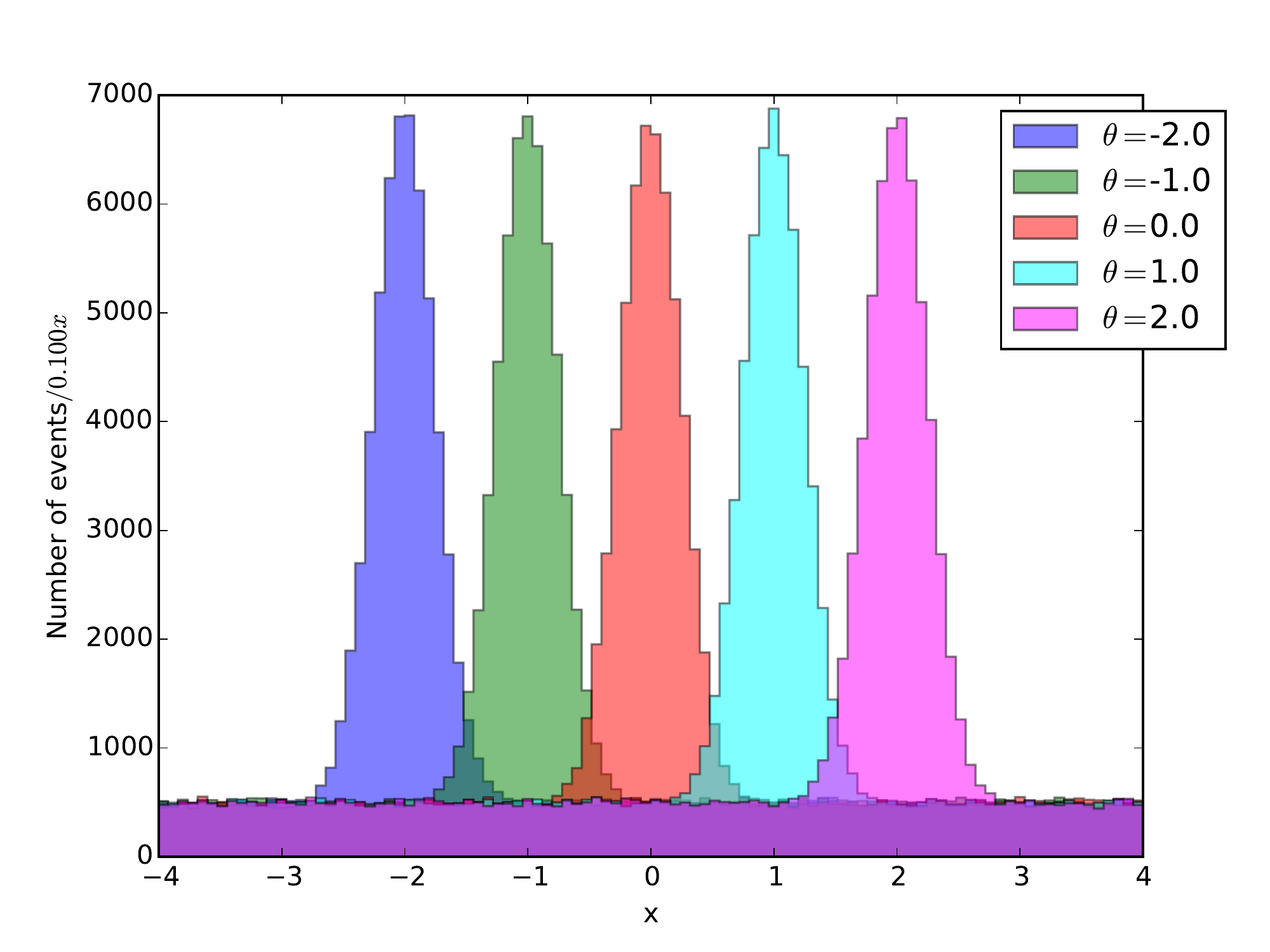}
\includegraphics[width=0.35\textwidth]{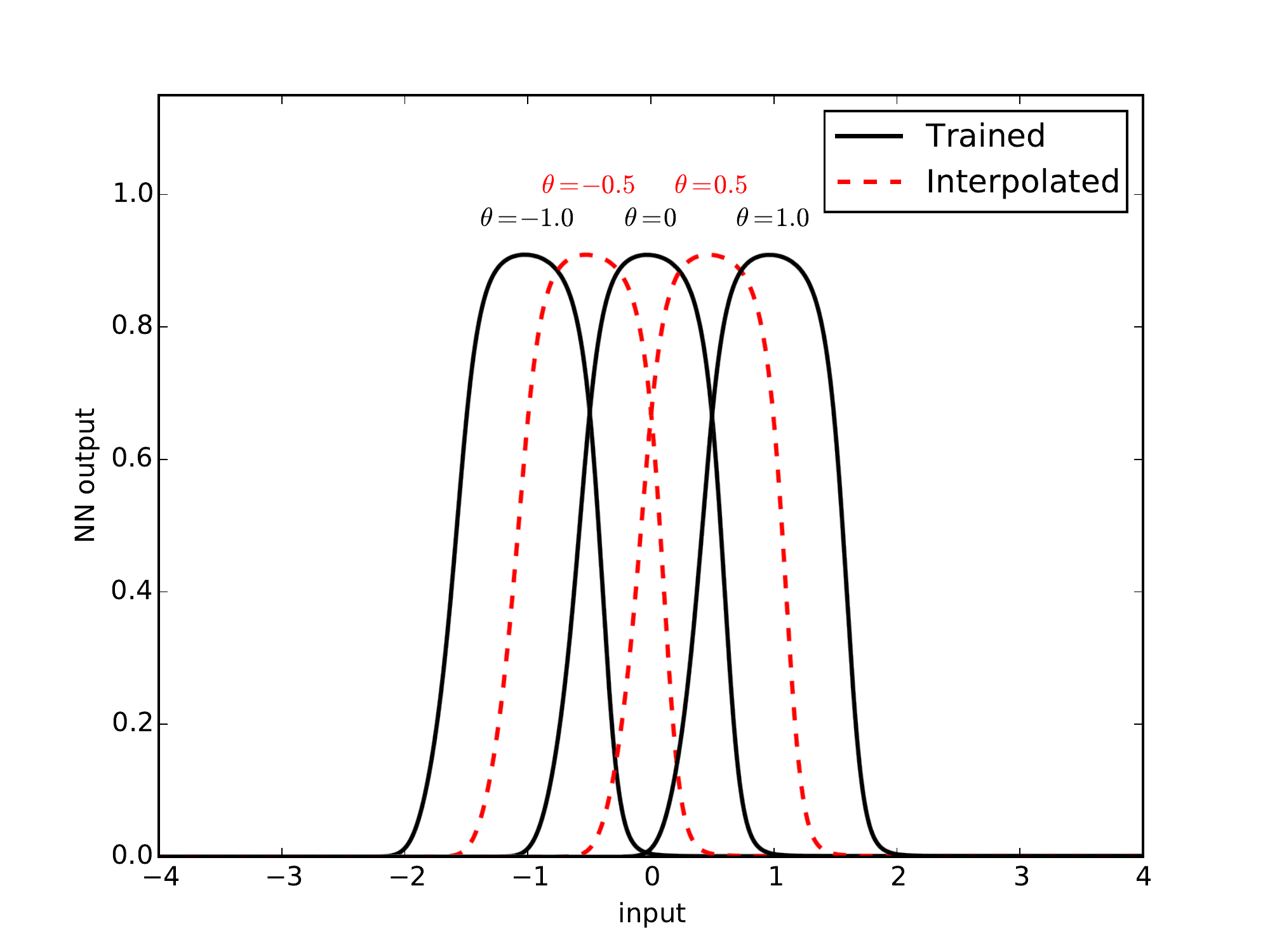}
\caption{Top, training samples in which the signal is drawn from a Gaussian and the background is uniform.
Bottom, neural network response as a function of the value of the input feature $x$, for various choices of the input parameter $\theta$; note that the single parameterized network has seen no training examples for $\theta=-1.5,-0.5,0.5,1.5$.}
\label{fig:toy}
\end{figure}

As a demonstration for a simple toy problem, we construct a parameterized network, which has a single input feature $x$ and a single parameter $\theta$. The network is trained using labeled examples where examples with label 0 are drawn from a uniform background and examples with label 1 are drawn from a Gaussian with mean $\theta$ and width $\sigma=0.25$. Training samples are generated with $\theta=-2,-1,0,1,2$; see Fig.~\ref{fig:toy}a.

As shown in Fig.~\ref{fig:toy}, this network generalizes the solution and provides reasonable output {\it even for values of the parameter where it was given no examples}. Note that the response function has the same shape for these values ($\theta=-1.5,-0.5,0.5,1.5$) as for values where training data was provided, indicating that the network has successfully parameterized the solution. The signal-background classification accuracy is as good for values where training data exist as it is for values where training data does not.

\section{1D Physical Example}

\begin{figure}
\includegraphics[width=0.2\textwidth]{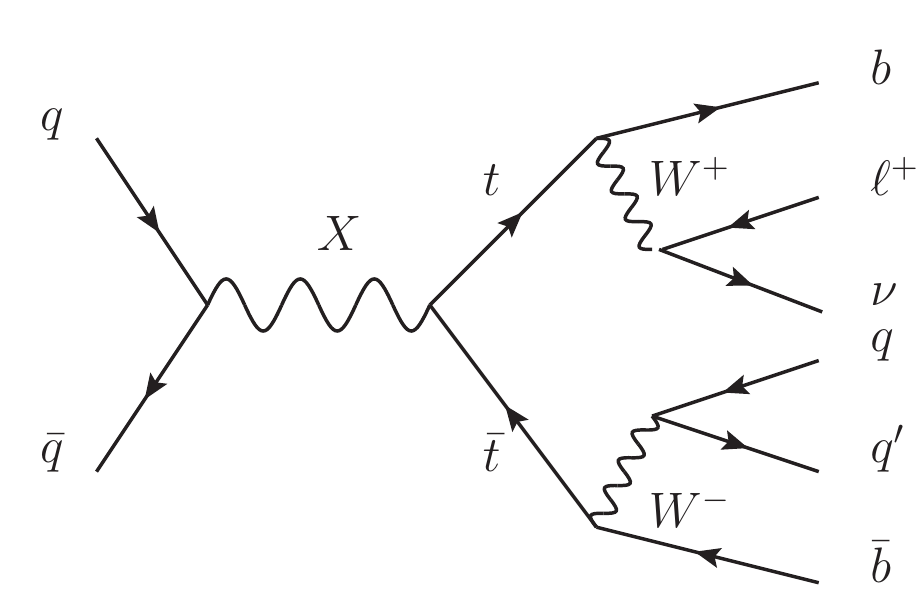}
\includegraphics[width=0.2\textwidth]{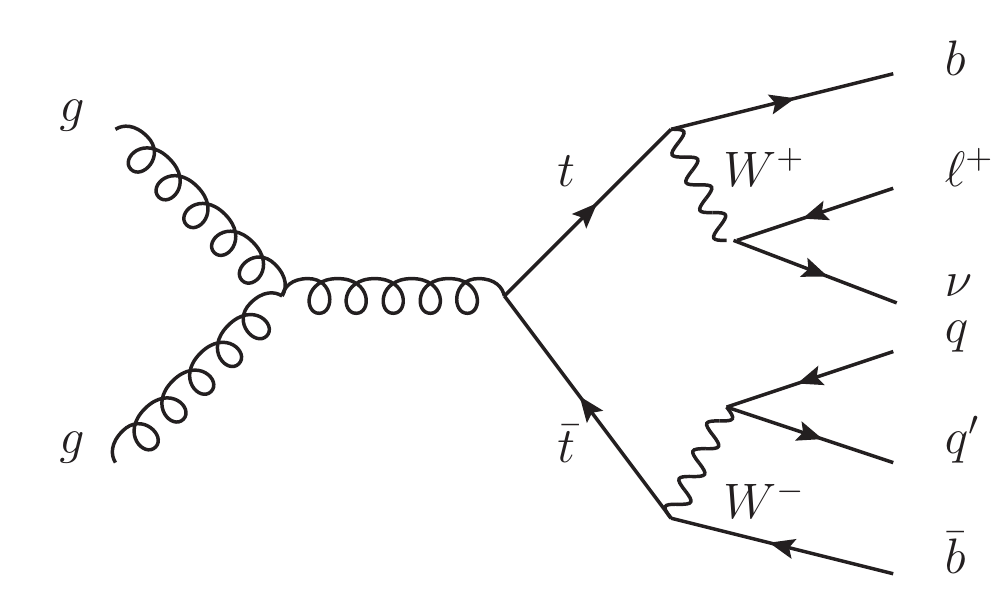}
\caption{ Feynman diagrams showing the production and decay of the hypothetical particle $X\rightarrow t\bar{t}$, as well as the dominant standard model background process of top quark pair production. In both cases, the $t\bar{t}$ pair decay to a single charged lepton ($\ell$), a neutrino ($\nu$) and several quarks ($q,b$).}
\label{fig:diag}
\end{figure}

A natural physical case is the application to the search for new particle of unknown mass. As an example, we consider the search for a new particle $X$ which decays to $t\bar{t}$.  We treat the most powerful decay mode, in which $t\bar{t}\rightarrow W^+bW^-\bar{b}\rightarrow qq'b\ell\nu \bar{b}$. The dominant background is standard model $t\bar{t}$ production, which is identical in final state but distinct in kinematics due to the lack of an intermediate resonance.   Figure~\ref{fig:diag} shows diagrams for both the signal and background processes.

\begin{figure}[h!]
\includegraphics[width=0.4\textwidth]{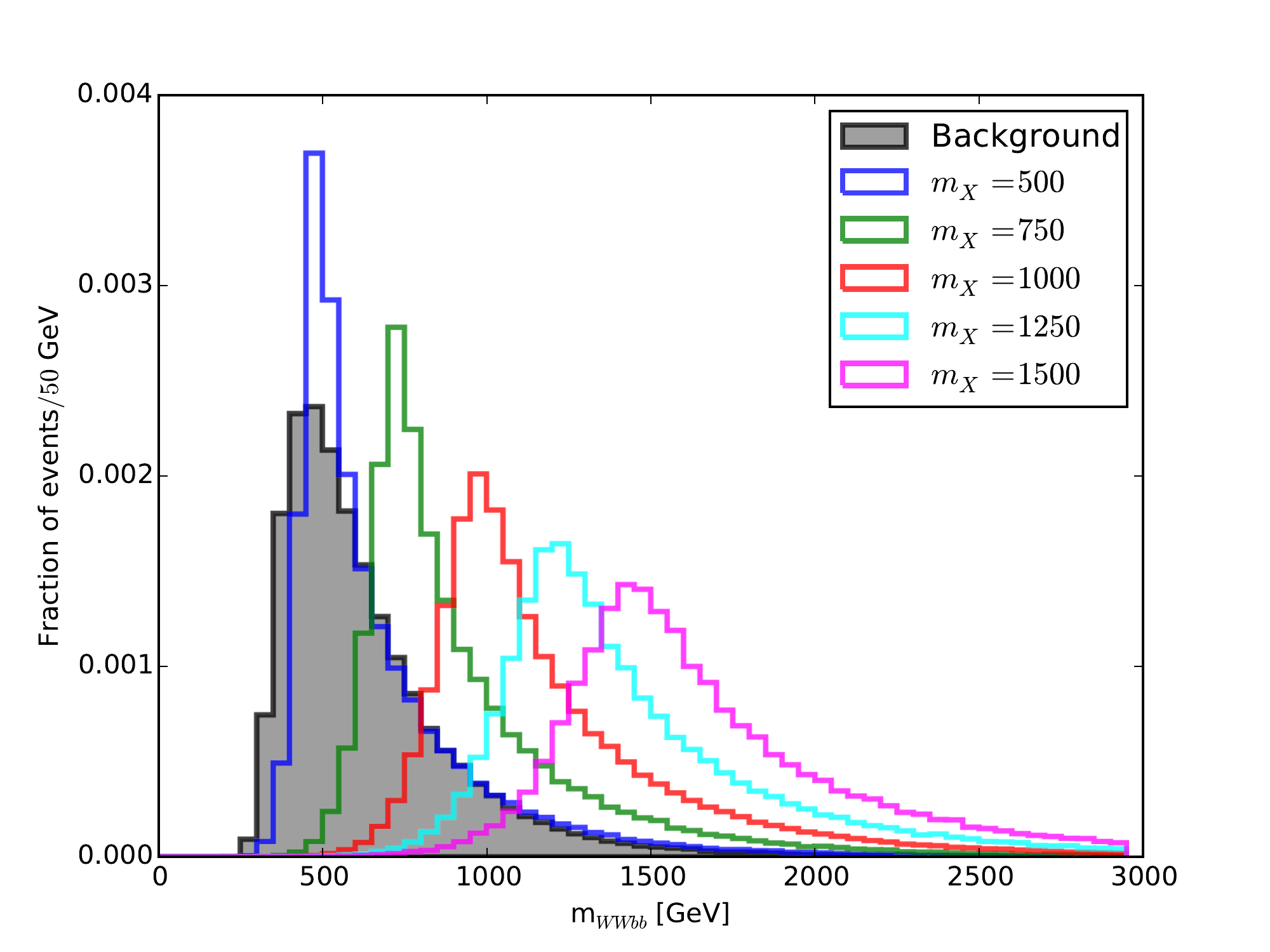}
\includegraphics[width=0.4\textwidth]{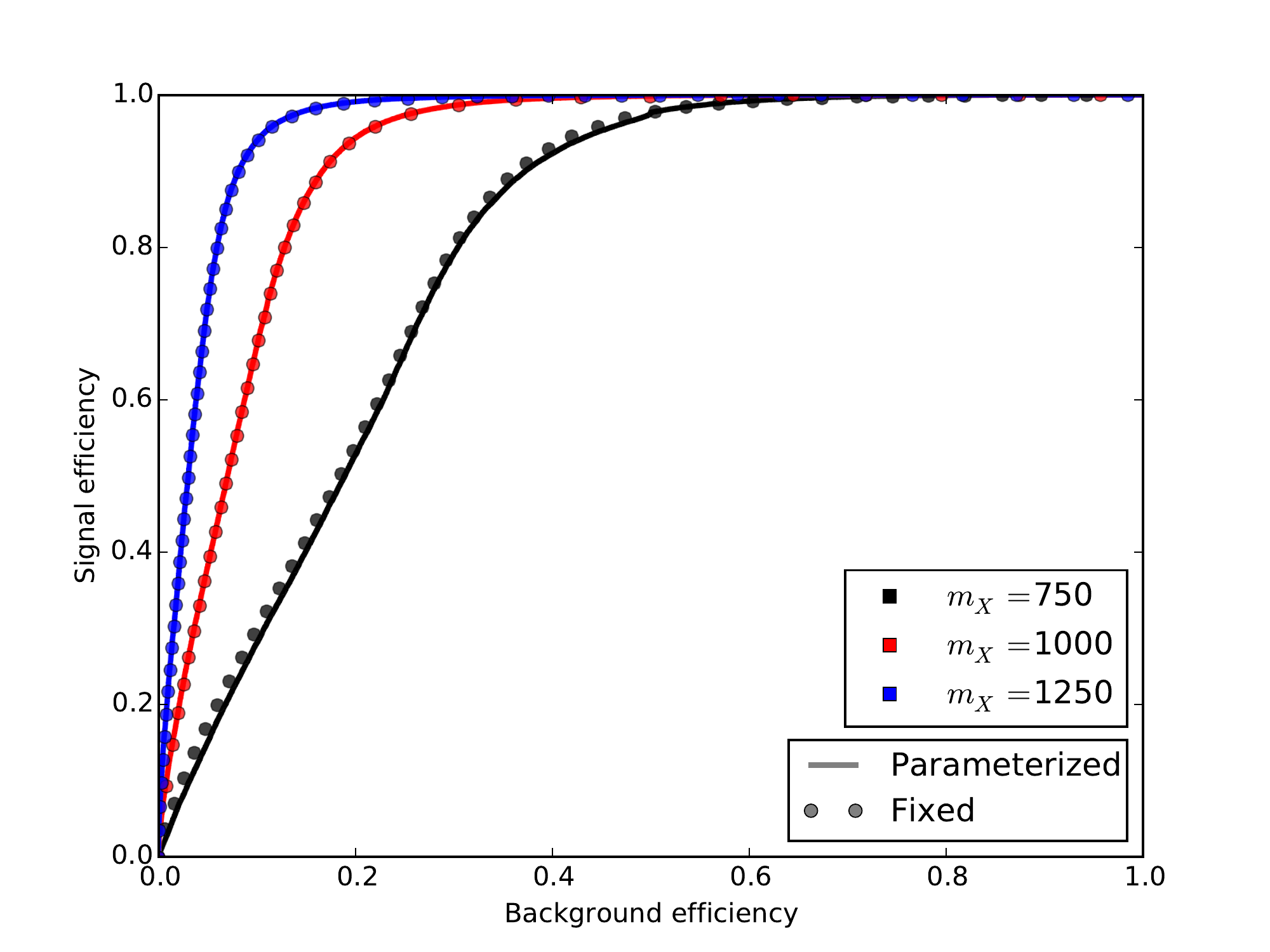}
\caption{ Top, distributions of neural network input $m_{WWbb}$ for the background and two signal cases.  Bottom, ROC curves for individual fixed networks as well as the parameterized network evaluated at the true mass, but trained only at other masses.  }
\label{fig:1dperf}
\end{figure}

We first explore the performance in a one-dimensional case. The single event-level feature of the network is $m_{WWbb}$, the reconstructed resonance mass, calculated using standard techniques identical to those described in Ref.~\cite{Aad:2013dza}.  Specifically, we assume resolved top quarks in each case, for simplicity. Events are are simulated at parton level with {\sc madgraph}5~\cite{Alwall:2011uj}, using {\sc pythia}~\cite{Sjostrand:2006za} for showering and hadronization and {\sc delphes}~\cite{deFavereau:2013fsa} with the ATLAS-style configuration for detector simulation.  Figure~\ref{fig:1dperf}a shows the distribution of reconstructed masses for the background process as well as several values of $m_X$, the mass of the hypothetical $X$ particle. Clearly the nature of the discrimination problem is distinct at each mass, though similar to those at other masses.

In a typical application of neural networks, one might consider various options:

\begin{itemize}
\item Train a single neural network at one intermediate value of the mass and use it for all other mass values as was done in Refs.~\cite{Aad:2014xea,Chatrchyan:2012yca}. This approach gives the best performance at the mass used in the training sample, but performance degrades at other masses.
\item Train a single neural network using an unlabeled  mixture of signal samples and use it for all other mass values. This approach may reduce the loss in performance away from the single mass value used in the previous approach, but it also degrades the performance near that mass point, as the signal is smeared.
\item Train a set of neural networks for a set of mass values as done in Refs.~\cite{Aaltonen:2012qt, Chatrchyan:2012tx}. This approach gives the best signal-background classification performance at each of the trained mass values, degrades for mass values away from the ones used in training, and leads to discontinuous performance as one switches between networks.
\end{itemize}

In contrast, we train a single neural network with an additional parameter, the true mass, as an input feature. For a learning task with $n$ event-level features and $m$ parameters, one can trivially reconcieve this as a learning task with $n+m$ features. Evaluating the network requires supplying the set of event-level features as well as the desired values of the parameters. 

These  neural networks are implemented using the multi-layer perceptron in PyLearn2, with outputs treated with a regressor method and logistic activation function. Input and output data are subject to preprocessing via a scikit-learn pipeline (i.e. MinMaxScaler transformation to inputs/outputs with a minimum and maximum of zero and one, respectively). Each neural network is trained with 3 hidden layers and using Nesterov's method for stochastic gradient descent. Learning rates were initiated at 0.01, learning momentum was set to 0.9, and minibatch size is set to treat each point individually (i.e. minibatch size of 1).  The training samples have approximately 100k examples per mass point.

The critical test is the signal-background classification performance. To measure the ability of the network to perform well at interpolated values of the parameter -- values at which it has seen no training data -- we compare the performance of a single fixed network trained at a specific value of $m_{X}^0$ to a parameterized network trained at the other available values other than $m_{X}^0$. For example, Fig.~\ref{fig:1dperf} compares a single network trained at $m_{X}^0=750$ GeV  to a parameterized network trained with data at $m_{X}=500,1000,1250,1500$ GeV.  The parameterized network's input parameter is set to the true value of the mass ($m_X^0$, and it is applied to data generated at that mass; recall that it saw no examples at this value of $m_X^0$ in training.  Its performance matches that of the single network trained at that value, validating the ability of the single parameterized network  to interpolate between mass values without any appreciable loss of performance.

\section{High-dimensional Physical Example}

The preceding examples serve to demonstrate the concept in one-dimensional cases where the variation of the output on both the parameters and features can be easily visualized. In this section, we demonstrate that the parameterization of the problem and the interpolation power that it provides can be achieved also in high-dimensional cases.

\begin{figure}
\includegraphics[width=0.2\textwidth]{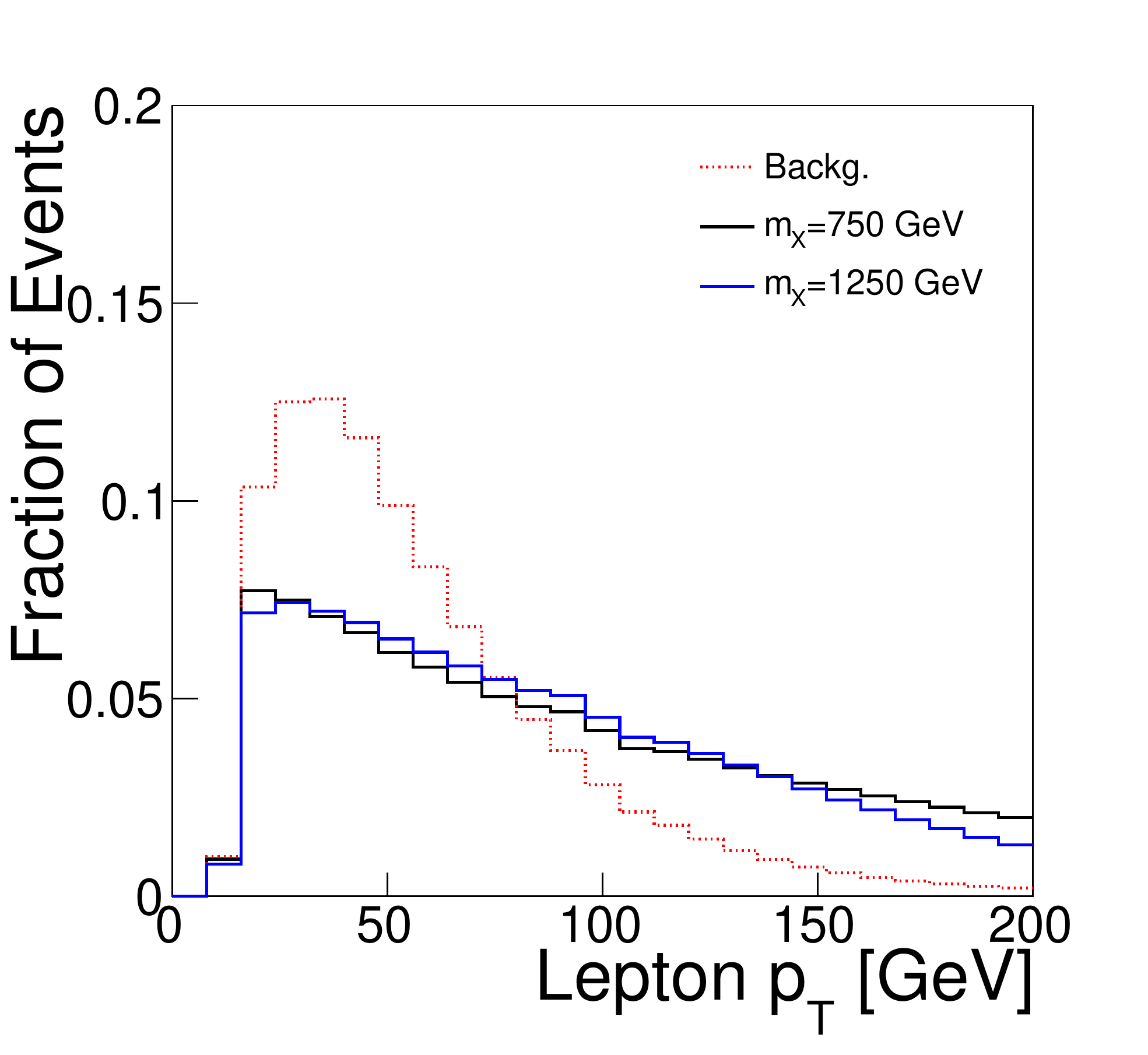}
\includegraphics[width=0.2\textwidth]{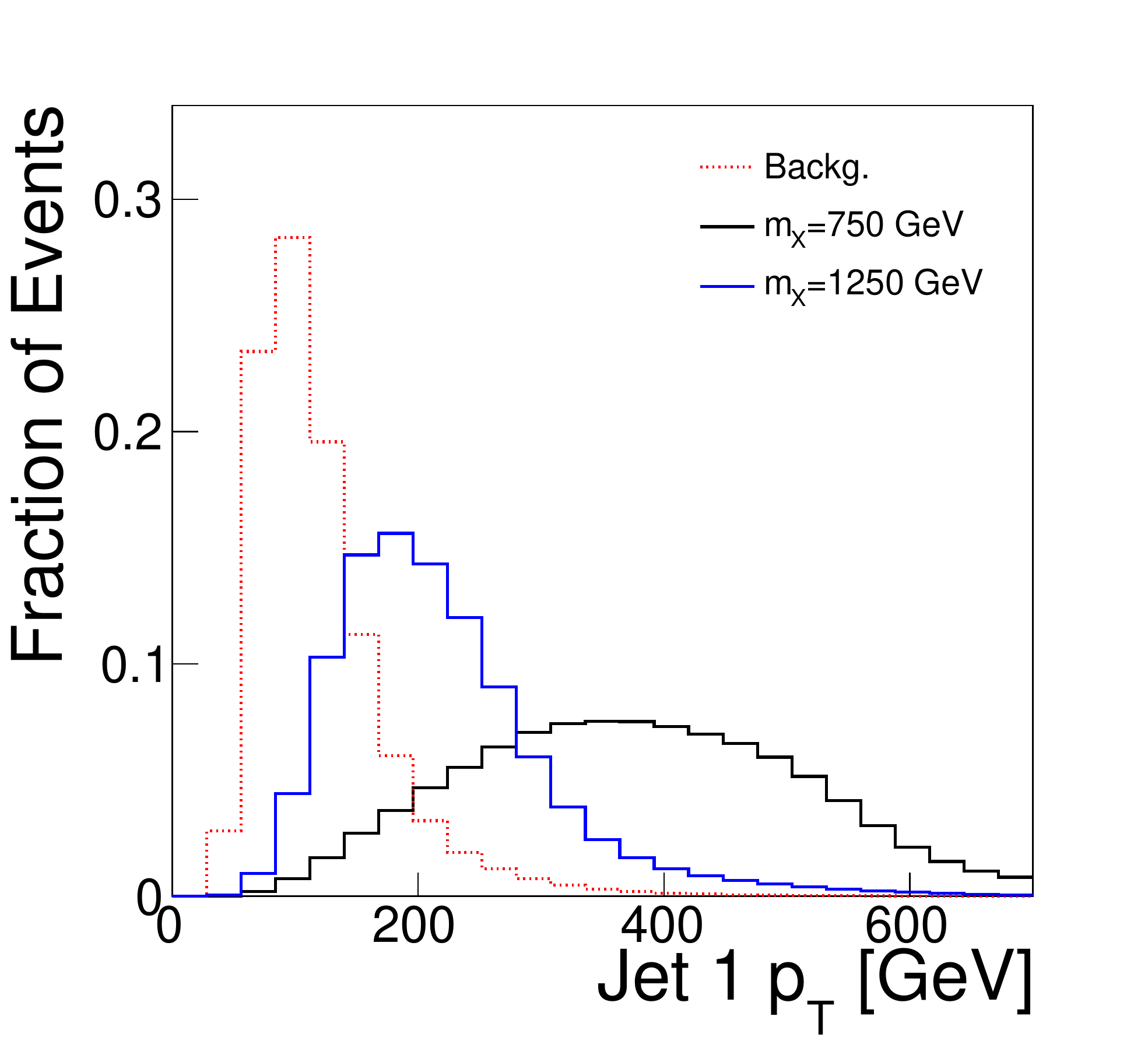}
\includegraphics[width=0.2\textwidth]{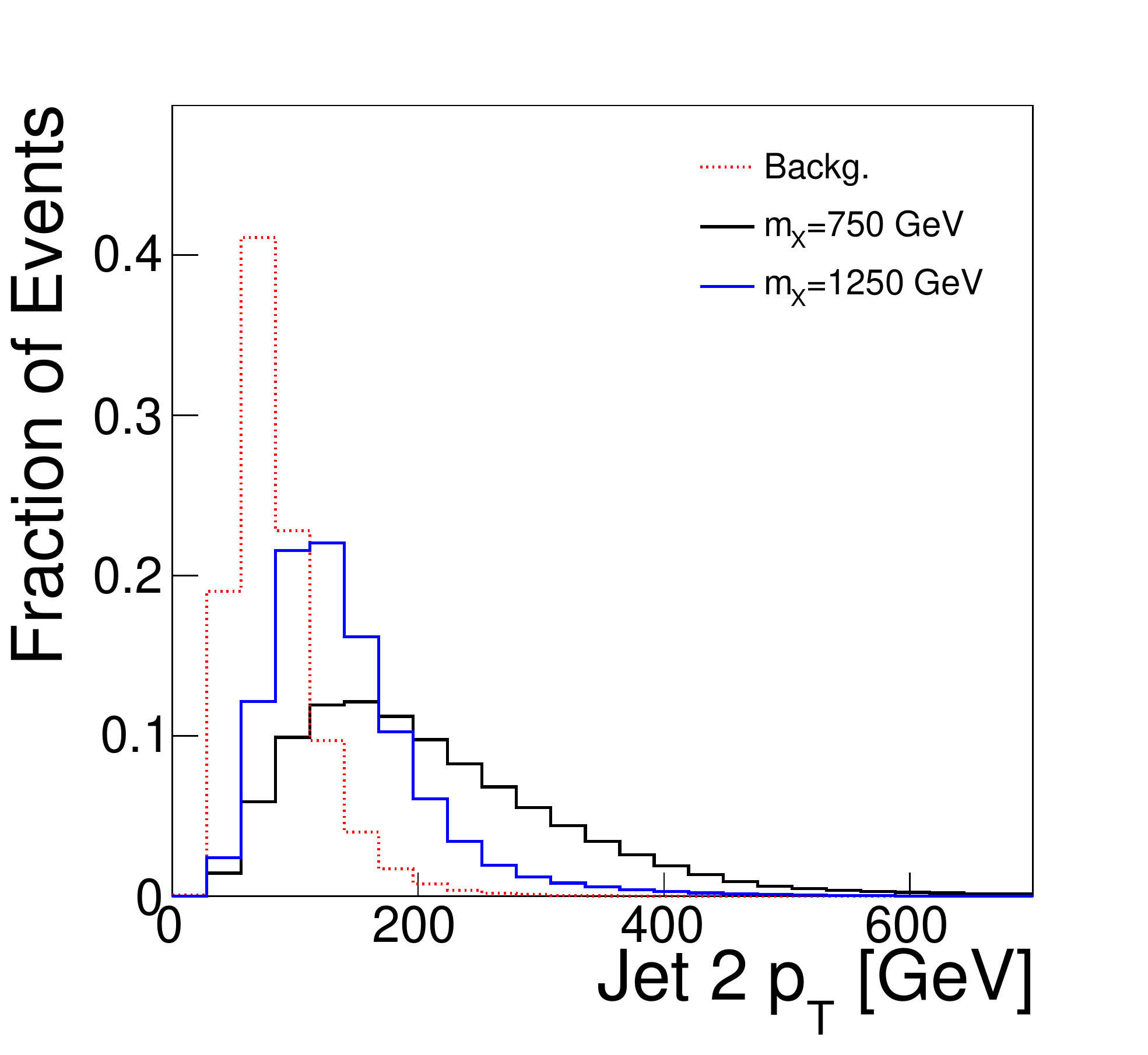}
\includegraphics[width=0.2\textwidth]{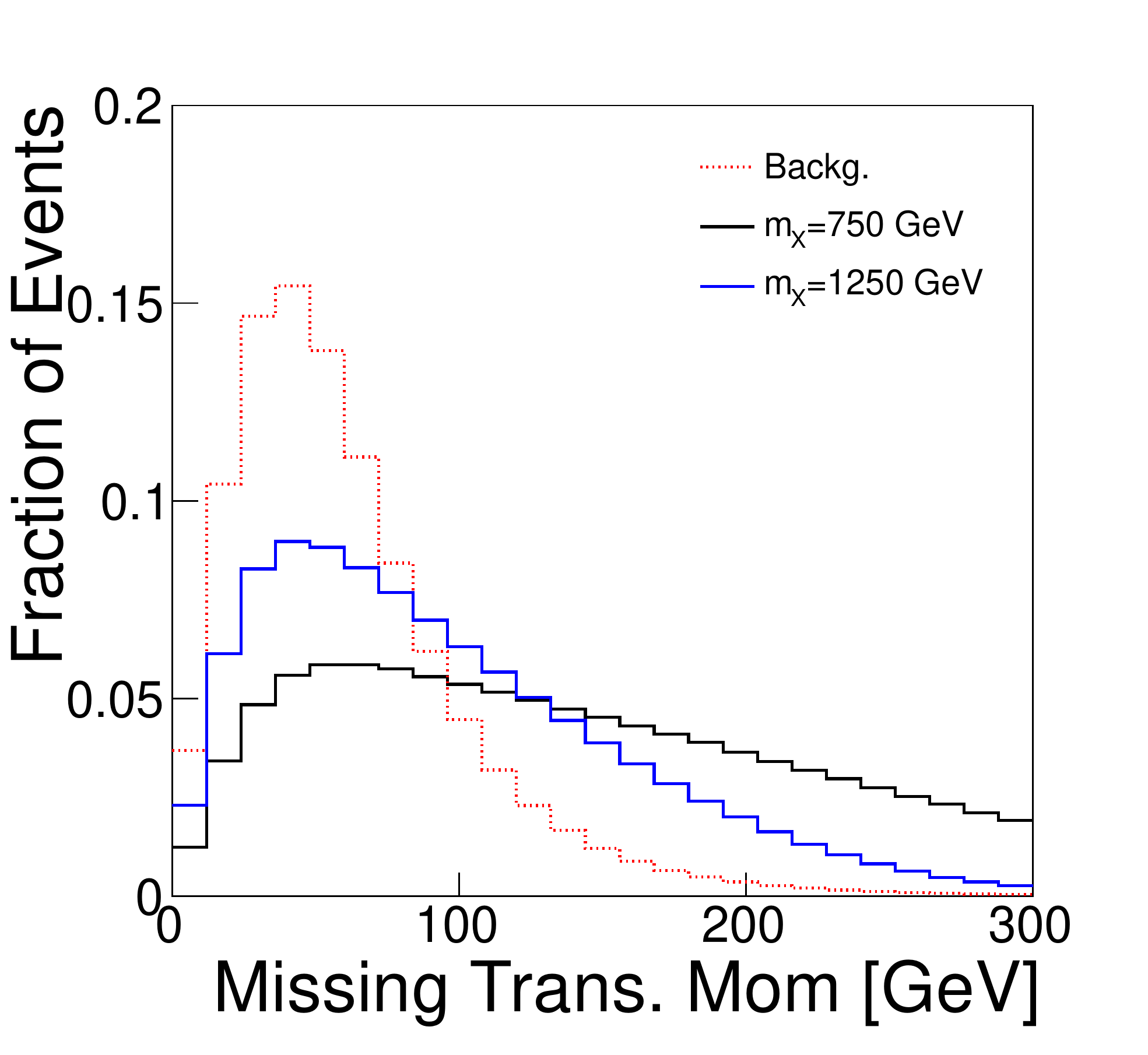}
\caption{ Distributions of some of the low-level event features for the decay of $X\rightarrow t\bar{t}$ with two choices of $m_X$ as well as the dominant background process.}
\label{fig:llvar}
\end{figure}

\begin{figure}
\includegraphics[width=0.2\textwidth]{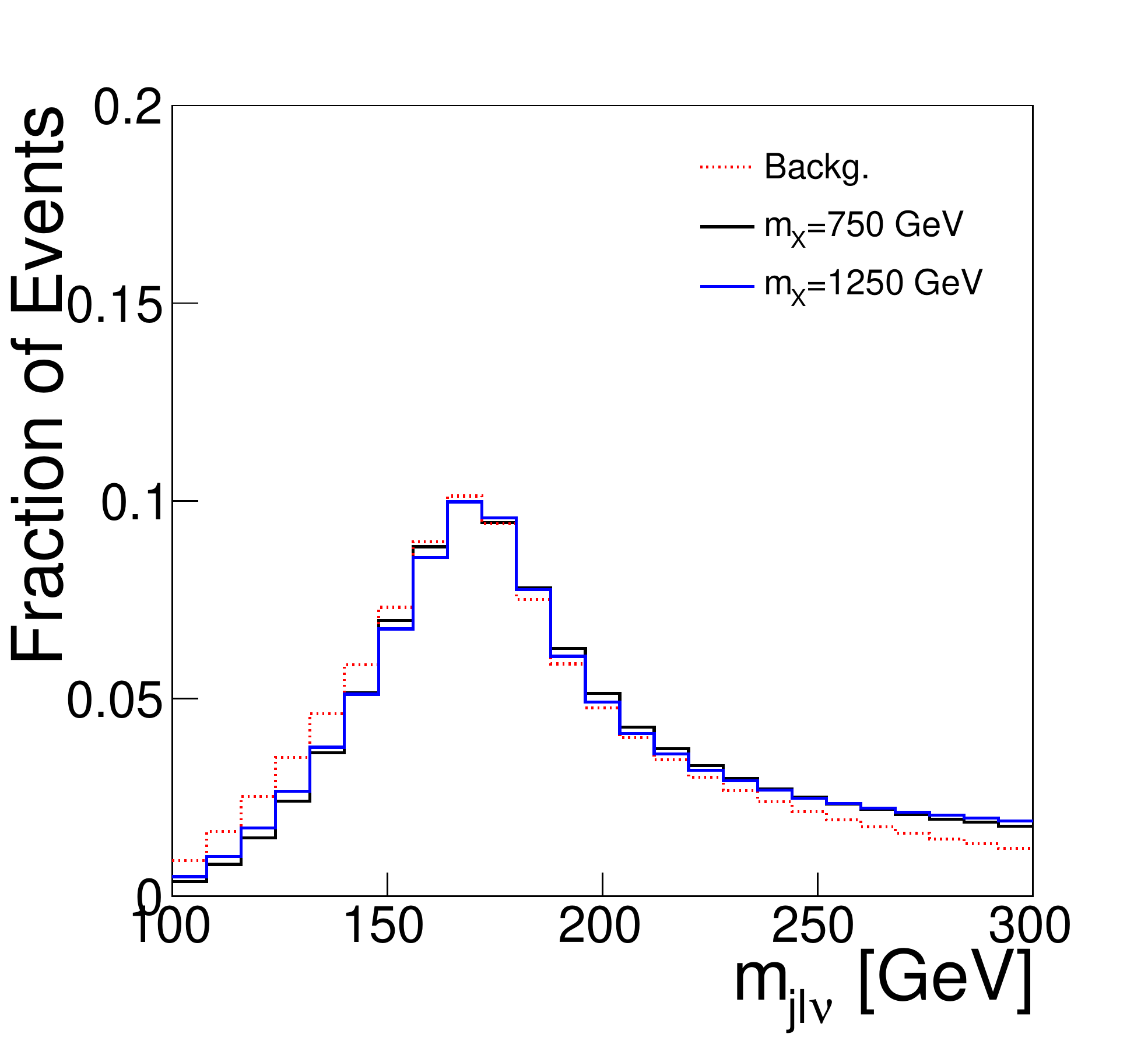}
\includegraphics[width=0.2\textwidth]{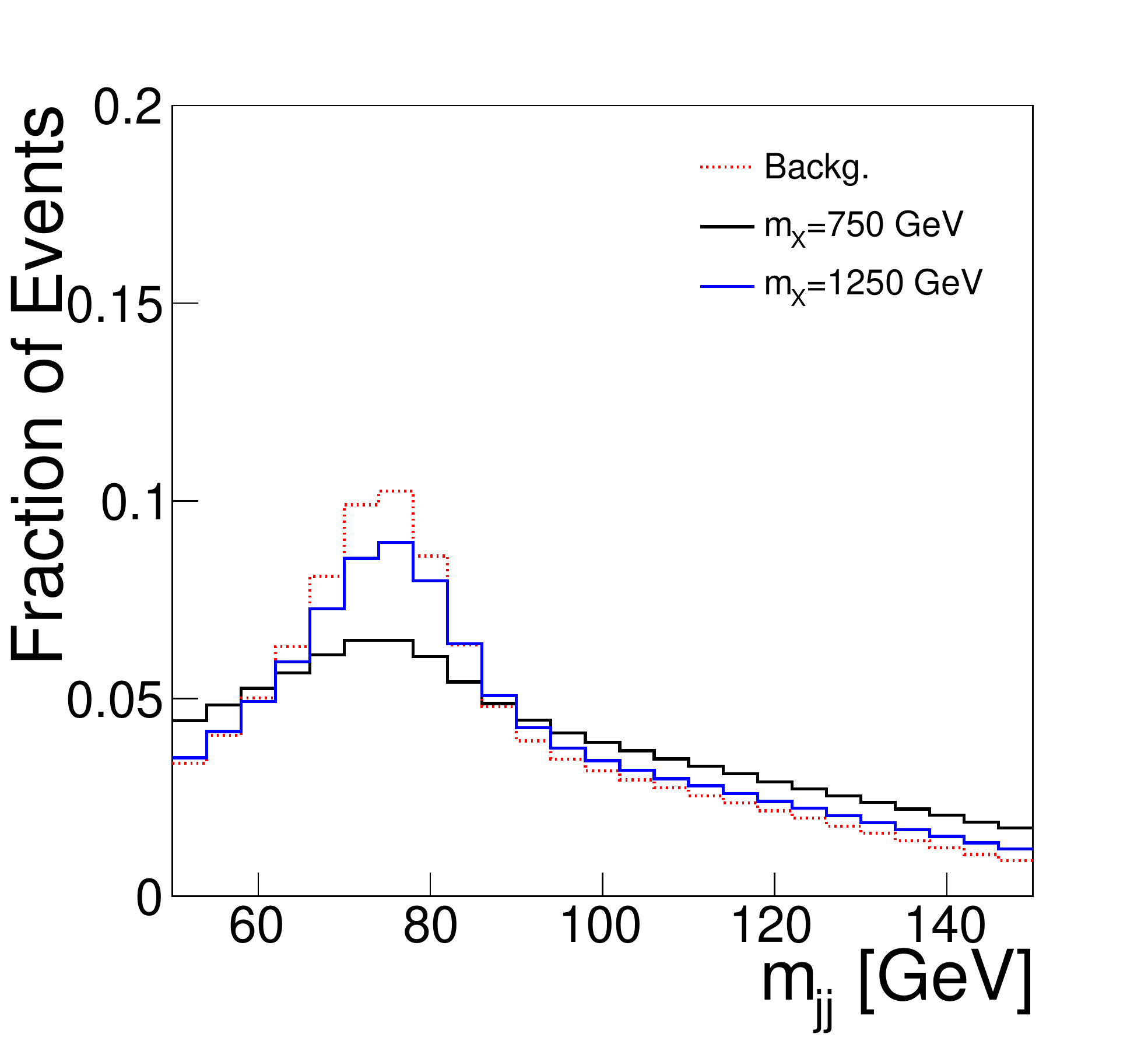}
\includegraphics[width=0.2\textwidth]{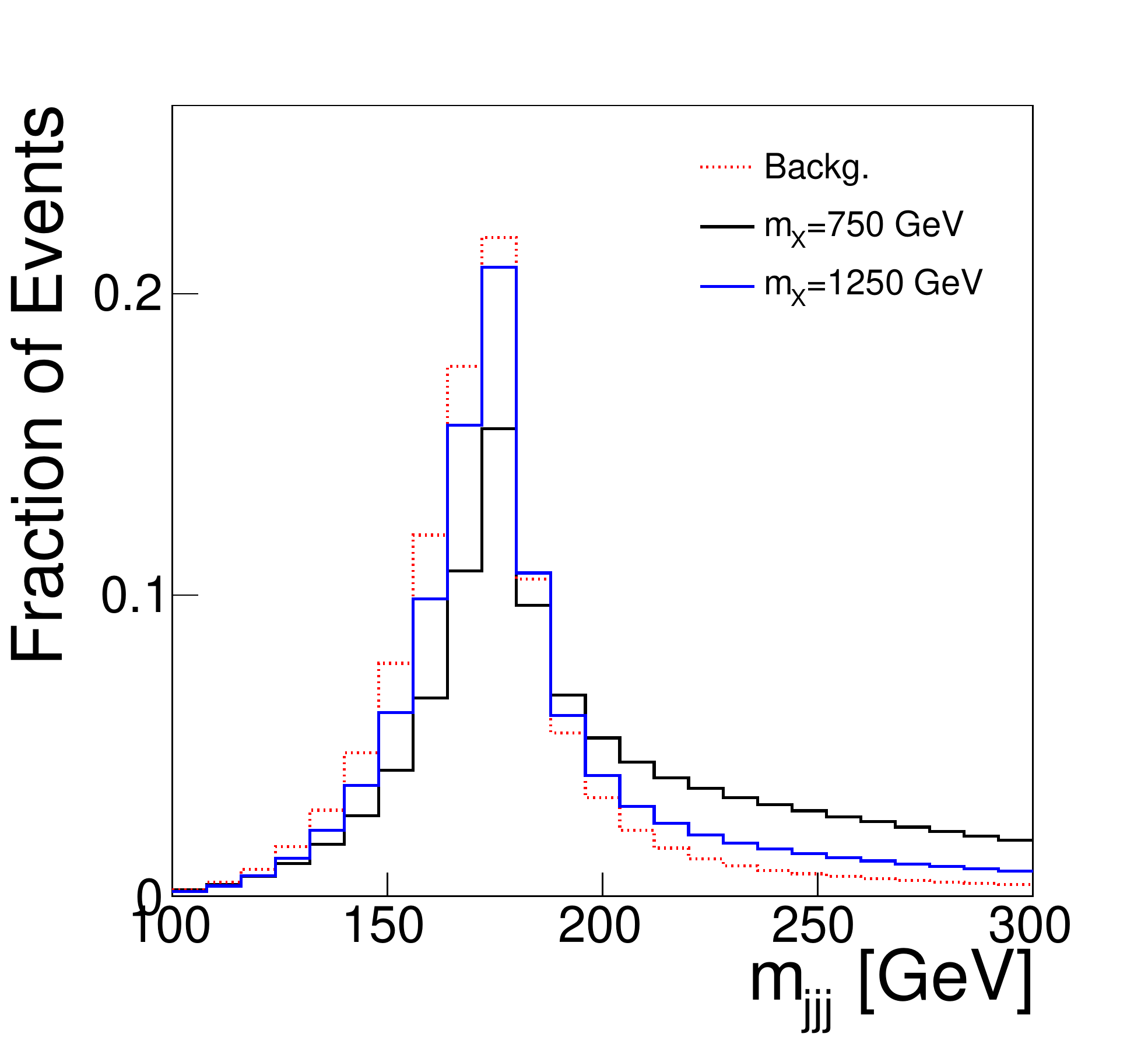}
\includegraphics[width=0.2\textwidth]{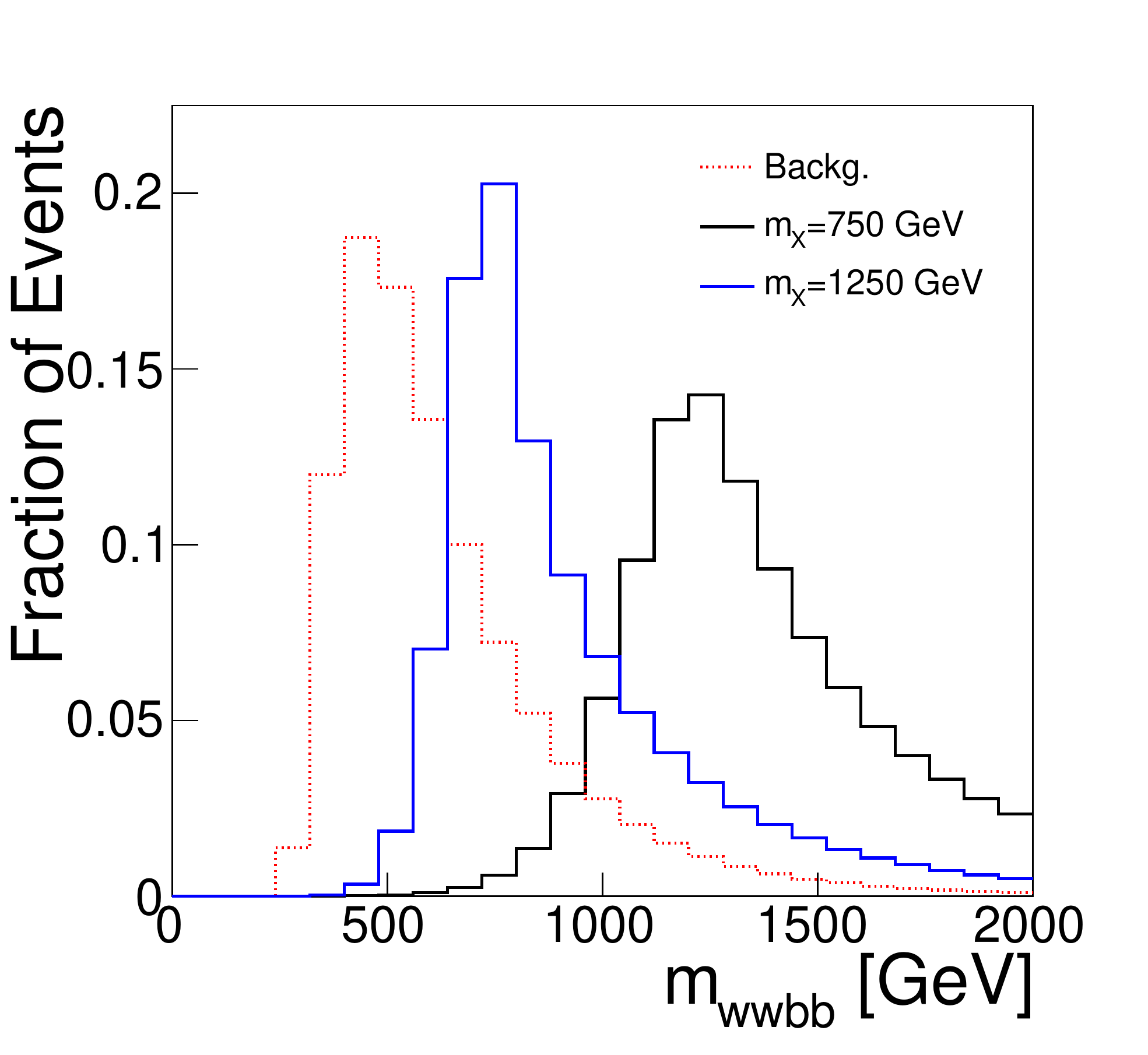}
\caption{Distributions of high-level event features for the decay of $X\rightarrow t\bar{t}$ with two choices of $m_X$ as well as the dominant background process; see text for definitions.}
\label{fig:hlvar}
\end{figure}

We consider the same hypothetical signal and background process as above, but now expand the set of features to include both low-level kinematic features which correspond to the result of standard reconstruction algorithms, and high-level features, which benefit from the application of physics domain knowledge. The low-level features are roughly the four-vectors of the reconstructed events, namely:

\begin{itemize}
\item the leading lepton momenta, 
\item the momenta of the four leading jets, 
\item the $b$-tagging information for each jets
\item the missing transverse momentum magnitude and angle
\end{itemize}

\noindent
for a total of 21 low-level features; see Fig~\ref{fig:llvar}.  The high-level features strictly combine the low-level information to form approximate values of the invariant masses of the intermediate objects. These are:

\begin{itemize}
\item the mass ($m_{\ell\nu}$) of the $W\rightarrow\ell\nu$, 
\item the mass ($m_{jj}$) of the $W\rightarrow qq'$, 
\item the mass ($m_{jjj}$) of the $t\rightarrow Wb\rightarrow bqq'$, 
\item the mass ($m_{j\ell\nu}$) of the $t\rightarrow Wb\rightarrow\ell\nu b$, 
\item the mass ($m_{WWbb}$) of the hypothetical $X\rightarrow t\bar{t}$,
\end{itemize}

\noindent
for a total of five high-level features; see Fig~\ref{fig:hlvar}.

The parameterized deep neural network models were trained on GPUs using the Blocks framework \cite{van_merrienboer_blocks_2015, bastien_theano:_2012, bergstra_theano:_2010}. Seven million examples were used for training and one million were used for testing, with 50\% background and 50\% signal. The architectures contain five hidden layers of 500 hidden rectified linear units with a logistic output unit. Parameters were initialized from  $\mathcal{N}(0,0.1)$, and updated using stochastic gradient descent with mini-batches of size 100 and 0.5 momentum. The learning rate was initialized to 0.1 and decayed by a factor of 0.89 every epoch. Training was stopped after 200 epochs.

\begin{figure}
\includegraphics[width=0.4\textwidth]{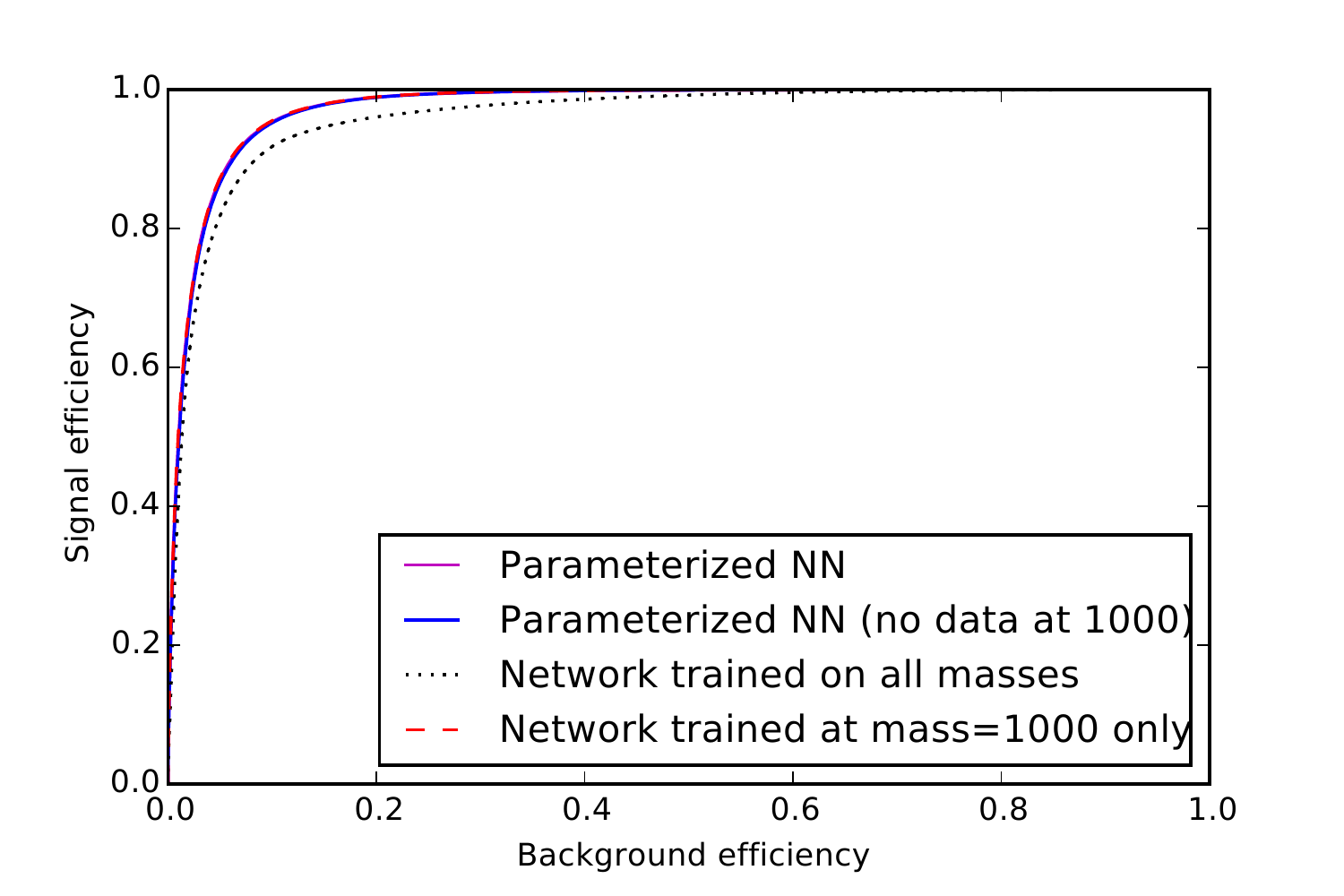}
\caption{ Comparison of the signal-to-background discrimination for four classes of networks for a testing sample with $m_X=1000$ GeV. A parameterized network trained on a set of masses ($m_X=500,750,1000,1250, 1500$) is compared to a single network trained at $m_X=1000$ GeV. The performance is equivalent. A second parameterized network is trained only with samples at $m_x=500,750,1250, 1500$, forcing it to interpolate the solution at $m_X =1000$ GeV. Lastly, a single non-parameterized network trained with all the mass points shows a reduced performance.  The results are indistinguishable for cases where the networks use only low-level features (shown) or low-level as well as high-level features (not shown, but identical).}
\label{fig:dnn_roc}
\end{figure}

The high dimensionality of this problem makes it difficult to visually explore the dependence of the neural network output on the parameter $m_{X}$. However, we can test the performance in signal-background classification tasks. We use three types of networks. A single parameterized network is trained using 7M training samples with masses $m_X=500,750,1000, 1250, 1500$ GeV and tested in a sample generated with $m_X=1000$ GeV; the performance is compared to a single fixed network trained with samples at $m_X=1000$ (with 7M training examples). In each case, we use approximately the same number of training and testing examples per mass point.  Fig~\ref{fig:dnn_roc} shows that the parameterized network matches the performance of the fixed network.  A more stringent follow-up test removes the $m_X=1000$ sample from the training set of the parameterized network, so that this network is required to interpolate its solution. The performance is unchanged, demonstrating that the parameterized network is capable of generalizing the solution even in a high-dimensional example.

\begin{figure}
\includegraphics[width=0.4\textwidth]{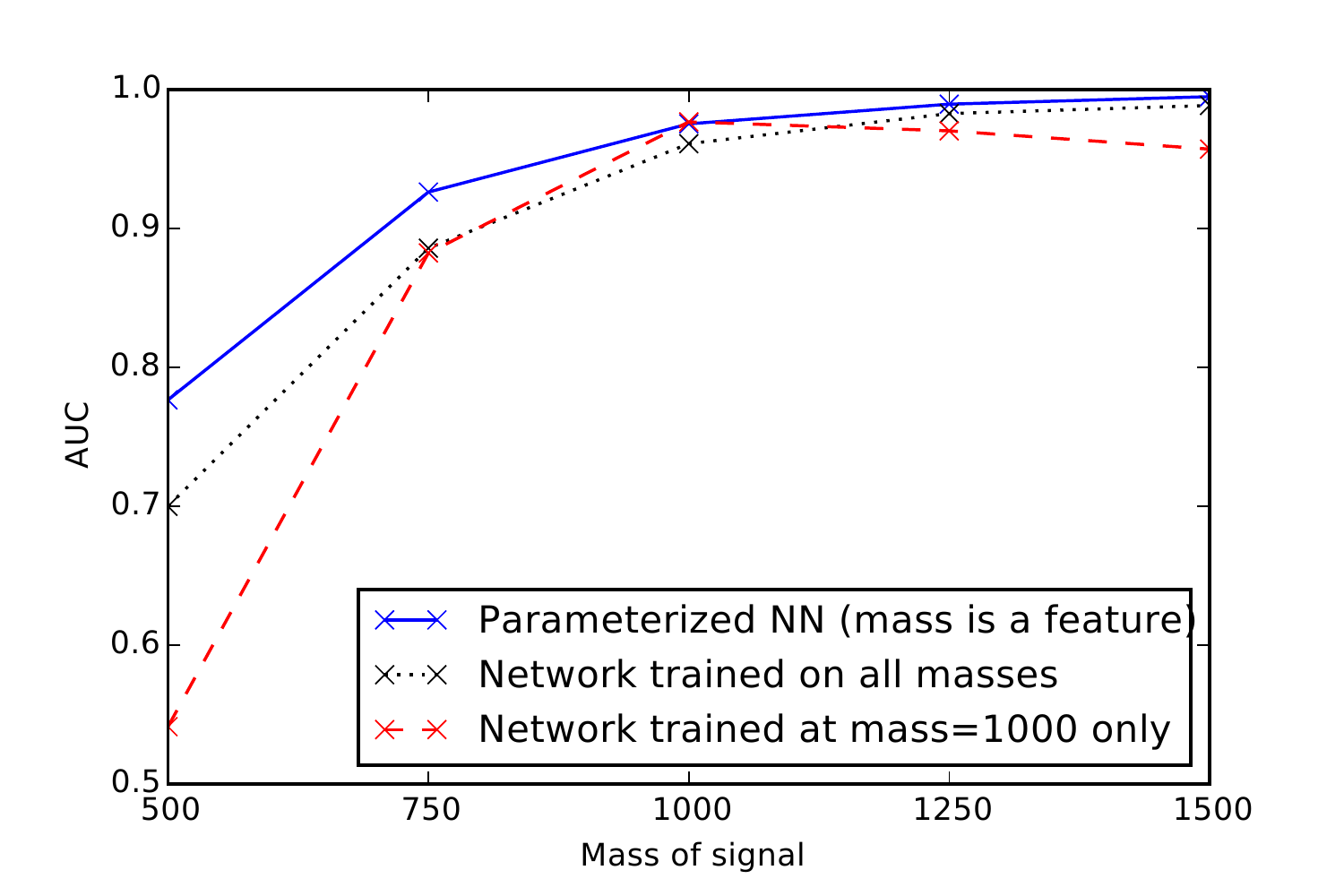}
\caption{ Comparison of the performance in the signal-background discrimination for the parameterized network, which learns the entire problem as a function of mass, and a single network trained only at $m_X=1000$ GeV.  As expected, the AUC score decreases for the single network as the mass deviates from the value in the training sample. The parameterized network shows improvement over this performance; the trend of improving AUC versus mass reflects the increasing separation between the signal and background samples with mass, see Figs.~\ref{fig:llvar} and~\ref{fig:hlvar}. For comparison, also shown in the performance a single network trained with an unlabeled mixture of signal samples at all masses.}
\label{fig:vmass}
\end{figure}

Conversely, Fig~\ref{fig:vmass} compares the performance of the parameterized network to a single network trained at $m_X=1000$ GeV when applied across the mass range of interest, which is a common application case. This demonstrates the loss of performance incurred by traditional approaches and recovered in this approach. Similarly, we see that a single network trained an unlabeled mixture of signal samples from all masses has reduced performance at each mass value tested.

In previous work, we have shown that deep networks such as these do not require the additional of high-level features~\cite{Baldi:2014kfa,Baldi:2014pta} but are capable of learning the necessary functions directly from the low-level four-vectors.  Here we extend that by repeating the study above without the use of the high-level features; see Fig~\ref{fig:dnn_roc}. Using only the low-level features, the parameterized deep network is achieves essentially indistinguishable performance for this particular problem and training sets of this size.

\section{Discussion}

We have presented a novel structure for neural networks that allows for a simplified and more powerful solution to a common use case in high-energy physics and demonstrated improved performance in a set of examples with increasing dimensionality for the input feature space.  While these example use a single parameter $\theta$, the technique is easily applied to higher dimensional parameter spaces.

Parameterized networks can also provide optimized performance as a function of nuisance parameters that describe systematic uncertainties, where typical networks are optimal only for a single specific value used during training. 
This allows statistical procedures that make use of profile likelihood ratio tests~\cite{Cowan:2010js}  to select the network corresponding to the profiled values of the nuisance parameters~\cite{cranmer2015}.

Datasets used in this paper containing millions of simulated collisions can be found in the UCI Machine Learning Repository~\cite{Bache+Lichman:2013} at \url{archive.ics.uci.edu/ml/datasets/HEPMASS}.

\subsection{Acknowledgements}
We thank Tobias Golling, Daniel Guest, Kevin Lannon, Juan Rojo, Gilles Louppe, and Chase Shimmin for useful discussions.  
KC is supported by the US National Science Foundation grants PHY-0955626, PHY-1205376, and ACI-1450310. KC is grateful to UC-Irvine for their hospitality while this research was initiated and the Moore and Sloan foundations for their generous support of the data science environment at NYU. We thank Yuzo Kanomata for computing support. We also wish to acknowledge a hardware grant from NVIDIA, NSF grant IIS-1550705, and a Google Faculty Research award to PB.
 
\bibliographystyle{unsrt}
\bibliography{paramnn}

\clearpage
\appendix

\end{document}